\documentclass[final,3p,times]{elsarticle}
\usepackage{dsfont}
\usepackage{epstopdf}

\usepackage{amssymb}

\newtheorem{defi}{Definition}[section]
\newtheorem{theorem}{Theorem}[section]
\newtheorem{lem}{Lemma}[section]
\newtheorem{propo}{Proposition}[section]
\newtheorem{corollary}{Corollary}[section]

\newtheorem{exmp}{Example}[section]
\newtheorem{remark}{Remark}[section]

\journal{arXiv}
\begin{document}

\begin{frontmatter}

\title{Laplacian integrality in $P_4$-sparse and $P_4$-extendible graphs}

\author[label1]{Renata R. Del-Vecchio}
\address[label1]{Federal Fluminense University, renata@vm.uff.br}
\author[label2]{\'Atila Arueira Jones}
\address[label2]{Federal Fluminense University, atilajones@id.uff.br}

\begin{abstract}
 Let $G$ be a simple graph and
$L = L(G)$ the Laplacian matrix of $G$. $G$ is called $L$-integral if all its Laplacian eigenvalues are integer numbers.
It is known that every cograph, a graph free of $P_4$, is $L$-integral. The class of $P_4$-sparse graphs and the class of $P_4$-extendible graphs contain the cographs. It seems natural to investigate if the graphs in these classes are still $L$-integral. In this paper we characterized the $L$-integral graphs for both cases, $P_4$-sparse graphs and $P_4$-extendible graphs.
\end{abstract}


\end{frontmatter}


\section{Introduction}
Let $G(V,E)$ be a simple graph on $n$ vertices, $D(G)=diag(d_{1}, \ldots,d_{n})$ the diagonal
matrix of its vertex degrees and $A(G)$, the \emph{adjacency} matrix of $G$.
Let $L(G) = D(G) - A(G)$ be the \emph{Laplacian}  matrix of $G$.
A graph $G$ is called $L$-integral when all eigenvalues of $L$ are integer numbers.
The search for Laplacian integral graphs has been done in special classes, as we can see in \cite{LAMA}, \cite{Grone Merris}
and \cite{MERRIS3}, for instance.

Although the study of integral graphs has come from a theoretical issue in the begining, recently this topic is associated to applications in physics and chemistry, as we can see in \cite{Multiprocessor}, \cite{Christandl} and \cite{Gutman}. In view of such applications, it becomes more important to completly characterize the integral graphs among  special classes of graphs.

It is well known that every cograph is Laplacian integral, \cite{MERRIS2}.
Cographs are graphs free of $P_4$
and a natural generalization of this class is the class of $P_4$-sparse graphs, \cite{Hoang}, graphs with
"few $P_4$", containing the cographs.

These graphs have been extensively studied because they have interesting structural properties that helped
in solving graph-theoretic problems (see \cite{Comeil}).

A question naturally posed in this context is if $P_4$-sparse graphs are Laplacian integrals.
In this article we answer negatively to this question, proving that there is no $P_4$-sparse graph with integer Laplacian eigenvalues, unless it is a cograph.\\

Another class of graphs, also based in the number of $P_4$'s as induced subgraphs, called $P_4$-extendible graphs, was introduced in \cite{Jamison2}. This class also contain the cographs and is different from the class of $P_4$-sparse graphs. We investigate the same question for this class, characterizing the $L$-integral graphs among them.

Besides this introduction we have three more sections. The second
one is devoted to the study of spider graphs and its spectrum,
an important tool to characterize $P_4$-sparse graphs. We also remember some basic notions and results required for what follows.
At the third section, we investigate the $P_4$-sparse graphs, presenting our main theorem and some examples.
Finally, in the fourth section, we prove an analogous result to the precedent case, characterizing the $L$-integral graphs within the class of $P_4$-extendible graphs as the cographs.

\section{\large{Basic notions and Spider graphs}}

\subsection{Laplacian spectrum}

The Laplacian spectrum of a graph $G$ consists of its $s$ distinct Laplacian eigenvalues and their multiplicities.
It will be denoted by
$$\xi(G)=\left(
            \begin{array}{cccc}
              \mu_1 & \mu_2 & \ldots & \mu_s \\
              r_1 & r_2 & \ldots & r_s \\
            \end{array}
          \right)
$$
where $\mu_i$ is a Laplacian eigenvalue of $G$ with multiplicity $r_i$, $i \in \{1,\ldots,s\}$. \\

We recall the following result, that will be used later:

\begin{propo}\cite{Mohar}\label{Mohar}
If $\overline{G}$ denotes the complement of the graph $G$ with $n$ vertices, then
$ \mu_{i+1}(\overline{G})= n-\mu_{n-i+1}(G)$, $i \in \{1,\ldots, n-1\}$ and $\mu_{1}(\overline{G})= 0$, considering $\mu_{i}$
displayed in a non increasing order. \\
\end{propo}

As an immediate consequence of this, we have that $G$ is $L$-integral if and only if $\overline{G}$ is $L$-integral.\\

\begin{remark}  \label{remark_l-espectral}
We also recall that the L-spectrum of the union of two graphs $G$ and $H$, is given by the
union of their L-spectra, $\xi(G \cup H) = \xi(G) \cup \xi(H)$.
Therefore, in order to have $G\cup H$  L-integral, it is necessary and sufficient that $G$ and $H$ are L-integral.\\
Consequently, for a disconnected graph $G$, we have that $G$ is $L$-integral if and only if each connected component is $L$-integral.
\end{remark}

\subsection{Spider graphs}

We now present the definition of a spider graph:
\begin{defi} \cite{Hoang}
$G(V,E)$ is a \textit{spider} if $V$ can be partitioned into sets
$S,C$ and $R$ such that:
\begin{itemize}
 \item $|S|=|C|\geqslant 2$
 \item $S=\{s_{1},\ldots,s_{k}\}$ is an independent set;
 \item $C=\{c_{1},\ldots,c_{k}\}$ is a clique;
 \item There are all edges between vertices of $R$ and $C$ and no edges between vertices of
$R$ and $S$.
\end{itemize}
The adjacence between the vertices of $S$ and $C$ is given by:
$s_{i}$ is adjacent to $c_{j}$ if and only if $i=j$ or else, $s_{i}$ is adjacent to $c_{j}$ if and only if $i\neq j$.
If the first case holds, the graph is called a \textit{thin spider}. In the other case the graph is called a \textit{thick spider}.

The set $C$ is the body of the spider, the set $S$ corresponds to the  spider's legs, and
the set $R$ is the spider's head. If $R$ is an empty set, the graph is called a headless spider.

\end{defi}

\noindent\textbf{Notation:} The thin spider will be denoted by $S_{t}[H,k,j]$, where the legs and the body have $k$ vertices each and
$H$ is the graph induced by the head, with $j$ vertices.
 Similarly, the thick spider will be represented by $S_T[H,k,j]$. If the thin  (respectively thick) spider is headless
 i.e., $R$ is empty, we will denote it  by $S_t[k]$ (respectively $S_T[k]$).\\

\begin{exmp} \label{exem_aranha}
Figure \ref{fig:19} shows a thin spider whose head is a graph $H$ (with three vertices) and a thick spider with a  head formed by the same graph $H$.

\begin{figure}[ht]
\centering
\includegraphics[scale=0.3]{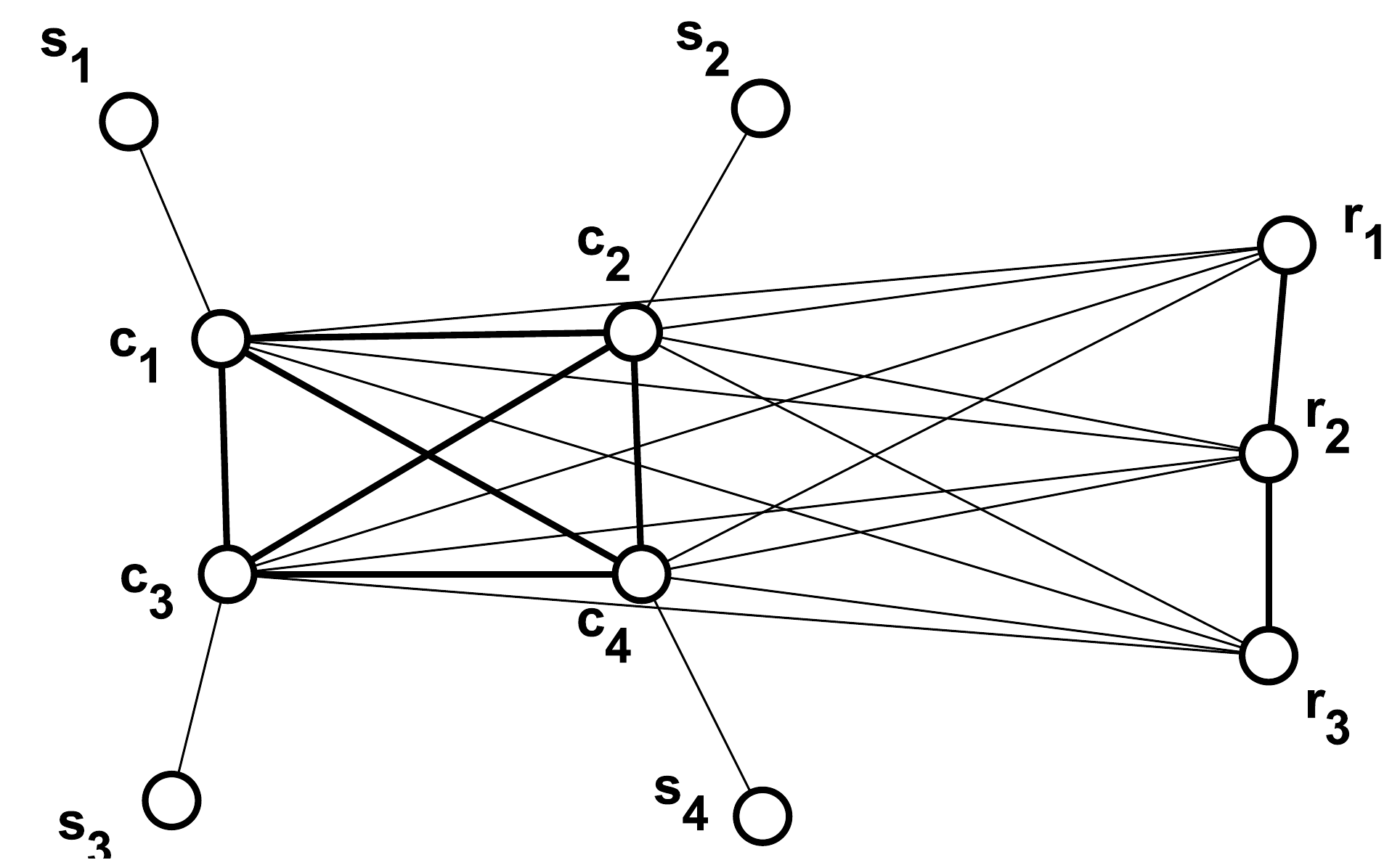} \quad
\includegraphics[scale=0.3]{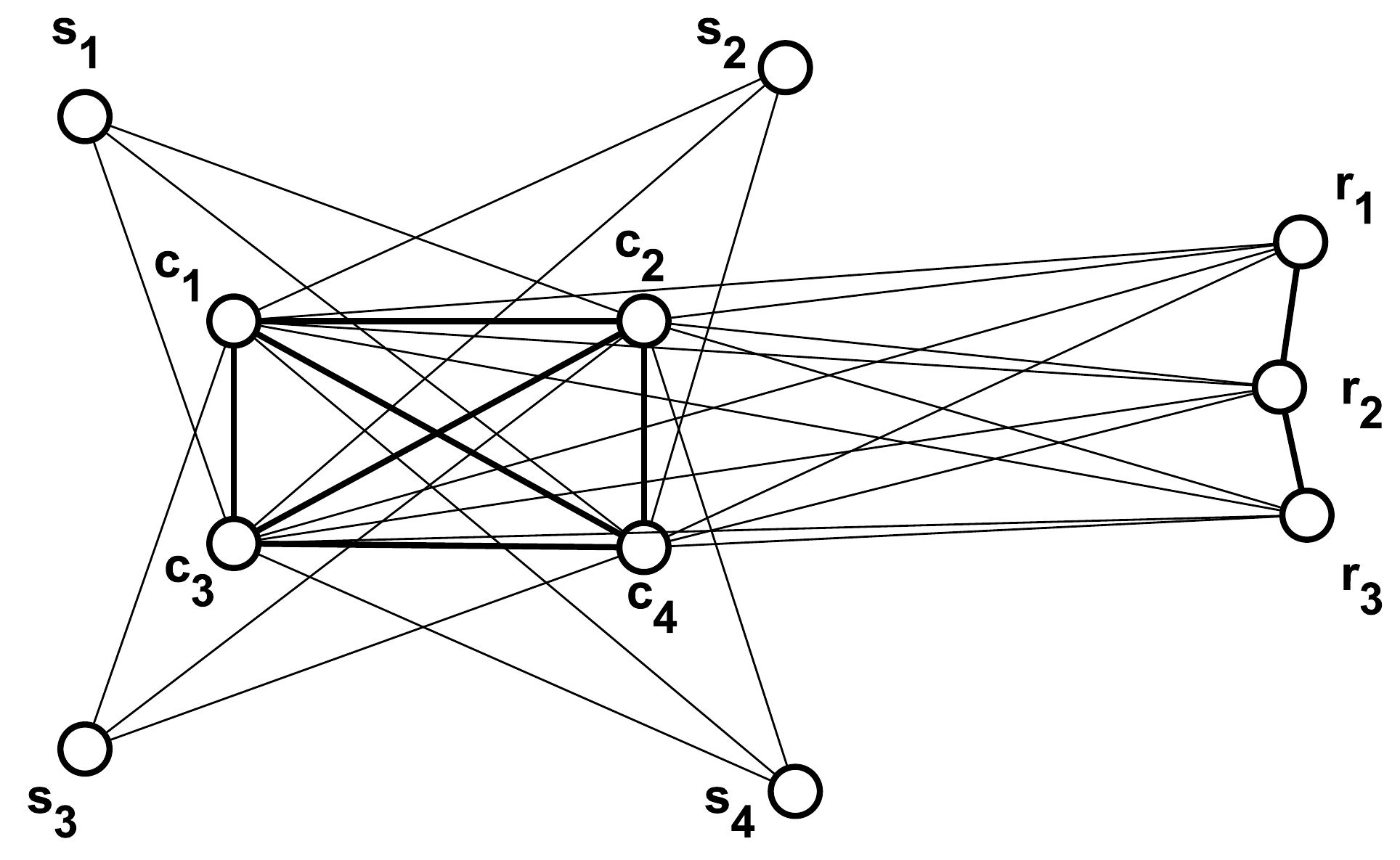}
\caption{$S_{t}[H,4,3]$ and $S_{T}[H,4,3]$}
\label{fig:19}
\end{figure}
\end{exmp}

\begin{remark} Every spider is a connected graph, even if the subgraph induced by its head is disconnected. Clearly, the complement of a spider is a spider. More specifically, given a thin spider with subgraph $H$ induced by the head, its complement is a thick spider with the same number of vertices in the body and the subgraph induced by the head is
$\overline{H}$ or, simply, $\overline{S_{t}[H,k,j]}=S_{T}[\overline{H},k,j]$.\\
\end{remark}

\begin{remark}
The path $P_4$ is a  headless spider whose body induces a subgraph isomorphic to $K_2$ and the complement of the path $P_4$ is isomorph to $P_4$.
\end{remark}

Henceforth, $\mathbf{1}_j$ and $\mathbf{0}_j$ are the vectors of order $j$ with all elements equal to $1$ and $0$, respectively, $\Theta_{j,\,k}$ denotes the $j \times  k$ all zeros matrix and $\mathbf{I}_j$ denotes the identity matrix of order $j$. Moreover, we denote the $j \times  k$ all ones matrix by $\mathds{J}_{j,\,k}$ and, in case of $k=j$, we simply denote it by $\mathds{J}_j$.

\begin{propo} \label{spect_magra}
Let $S_{t}[H,k,j]$ be a thin spider where $H$ is an empty subgraph (a graph without edges). Then its Laplacian spectrum is:

$$\xi(S_{t}[H,k,j])=\left(
                      \begin{array}{ccccc}
                        \frac{k+j+2 \pm \sqrt{(k+j)^2+4}}{2} & k & \frac{k+j+2 \pm \sqrt{(k+j)^2+4-4k}}{2} & 0\\
                        & & & \\
                        k-1,k-1 & j-1 & 1,1 & 0\\
                      \end{array}
                    \right)
$$
If it is a headless thin spider, $S_{t}[k]$, its Laplacian spectrum is:
$$\xi(S_{t}[k])=\left(
                      \begin{array}{ccc}
                        \frac{k+2 \pm \sqrt{k^2+4}}{2} & 0 & 2\\
                        & & \\
                        k-1,k-1 & 1 & 1\\
                      \end{array}
                    \right)
                    $$
\end{propo}

This notation means that $\frac{k+j+2 + \sqrt{(k+j)^2+4}}{2}$ is an eigenvalue with multiplicity $k-1$ and
$\frac{k+j+2 - \sqrt{(k+j)^2+4}}{2}$ is an eigenvalue with multiplicity $k-1$.
The same applies for the other cases where it appears $\pm$.\\

\noindent \textbf{Proof:}
Let $S_{t}[H,k,j]$ be a thin spider graph, whose head is an independent set ($H$ is a graph without edges with at least one vertex).
For simplicity we write $S_{t}$ instead of $S_{t}[H,k,j]$.
We label the vertices of the spider so that the matrix
$L(S_t)$ is written in blocks, expressing the links between body, legs and head.

$$\left(
  \begin{array}{ccc}
   \hbox{body x body} & \hbox{body x leg} & \hbox{body x head} \\
    \hbox{leg x body} & \hbox{leg x leg} & \hbox{leg x head}  \\
    \hbox{head x body} & \hbox{head x leg} & \hbox{head x head} \\
  \end{array}
\right)$$
The degrees  of the vertices of the  body, vertices of the head and vertices of legs are  $k+j$,  $k$ and $1$, respectively.
Then the matrix $L(S_t)=L$ can be written as a block matrix:


$$L(S_{t})=
\left[
  \begin{array}{ccc}
    [(k+j+1)\mathds{I}-\mathds{J}]_{k,\, k} & -\mathds{I}_{k} & -\mathds{J}_{k,\, j} \\
    -\mathds{I}_{k} & \mathds{I}_{k} & \Theta_{k ,\, j} \\
    -\mathds{J}_{j,\, k} & \Theta_{j,\, k} & k.\mathds{I}_{j} \\
  \end{array}
\right],
$$

Note that, for each  block matrix composing the matrix $L(S_{t})$,
the sum of its rows have the same value, leading to an equitable partition, $S$, $C$ and $R$.
Therefore, we can consider the matrix $L'_{3\times 3}$, whose entries are
such sums:

$$L'=\left(
            \begin{array}{ccc}
              j+1 & -1 & -j \\
              -1 & 1 & 0 \\
              -k & 0 & k \\
            \end{array}
          \right)
$$
Then, a known result about equitable partitions (see \cite{Brower})  ensures that the eigenvalues of
$L'$, which can be easily obtained, are also eigenvalues of $L(S_{t})$.


$$ x=0  ~\hbox{and}  ~\ x=\frac{k+2+j\pm\sqrt{(k+j)^{2}+4-4k}}{2}, $$
are the eigenvalues of $L'$ and then are also eigenvalues of $L$.

On the other hand, for each $u_j \in \mathds{R}^{j}$ orthogonal to ${\textbf{1}}_{j}$, if
$\textbf{v}=\left[
    \begin{array}{c}
    {\textbf{0}}_{k} \\
    {\textbf{0}}_{k}\\
    \textbf{u}_j\\
    \end{array}
    \right] \in \mathds{R}^{2k+j}$
then $L.\textbf{v}=k.\textbf{v}$. Hence, $k$ is an eigenvalue of the matrix $L$ corresponding to the eigenvector $\textbf{v} \in \mathds{R}^{2k+j}$. As there are $j-1$ linearly independent vectors in $\mathds{R}^{j}$ orthogonal to ${\textbf{1}}_{j}$, then $k$ is an laplacian eigenvalue of $S_{t}$ with multiplicity at least $j-1$. If the head of the spider has only one vertex ($j=1$) then the vector $\textbf{u}$ does not exist, so $k$ isn`t an eigenvalue.

We will prove now that $\frac{k+j+2 \pm \sqrt{(k+j)^2+4}}{2}$  is another eigenvalue of this graph.
By the definition of spider, its body must have at least two vertices ($k\geq 2$), so for each vector $\textbf{u} \in \mathds{R}^{k}$ orthogonal to ${\textbf{1}}_{k}$, consider the vector
$\textbf{w}= \left[
   \begin{array}{c}
     \textbf{u} \\
       \frac{k+j+ \sqrt{(k+j)^2+4}}{2}\textbf{u}\\
         {\textbf{0}}_{j} \\
          \end{array}
          \right] \in \mathds{R}^{2k+j}$.

For simplicity, set $p=k+j$ e $q=(k+j)^2+4$. Then\\
$L.\textbf{w}=L.\left[
        \begin{array}{c}
         \textbf{u} \\
         \frac{p+\sqrt{q}}{2}\textbf{u}\\
          {\textbf{0}}_{j} \\
           \end{array}
          \right]=
\left[
               \begin{array}{c}
                 \frac{p+2-\sqrt{q}}{2}\textbf{u}\\
                 \frac{p-2+\sqrt{q}}{2}\textbf{u} \\
                 {\textbf{0}}_{j} \\
               \end{array}
             \right]=
\frac{p+2-\sqrt{q}}{2}\left[
                           \begin{array}{c}
                             \textbf{u} \\
                             \frac{p+\sqrt{q}}{2}\textbf{u}\\
                              {\textbf{0}}_{j} \\
                               \end{array}
                         \right]=\frac{p+2-\sqrt{q}}{2}\textbf{w}$

Therefore $\frac{p+2-\sqrt{q}}{2}$ is an $L$-eigenvalue of the graph with multiplicity at least $k-1$.

By similar procedure, we can conclude that
$\frac{p+2+\sqrt{q}}{2}$ is an $L$-eigenvalue with multiplicity at least $k-1$, for the eigenvector
$\textbf{w}= \left[
                             \begin{array}{c}
                             \textbf{u} \\
                             \frac{p-\sqrt{q}}{2}\textbf{u}\\
                             {\textbf{0}}_{j} \\
                             \end{array}
                            \right] \in \mathds{R}^{2k+j}$.

As we have obtained exactly $j+2k$ $L$-eigenvalues, which is the order of the graph $S_t$, the proof is completed for this case. \\

If the spider $S_t$ is headless, then its Laplacian matrix is given by:
$$L(S_{t})=
\left[
  \begin{array}{cc}
    [(k+1)\mathds{I}-\mathds{J}]_{k,\, k} & -\mathds{I}_{k} \\
    -\mathds{I}_{k} & \mathds{I}_{k}  \\
  \end{array}
\right]
$$
Proceeding analogously  to the previous case, considering $j =0 $ when convenient, we obtain what we wanted.
\begin{flushright}
$\blacksquare$
\end{flushright}

From the proof of the proposition above, we can state the following theorem:

\begin{theorem} \label{integral magra}
If $G$ is a thin spider then $G$ is not L-integral.
\end{theorem}

\noindent \textbf{Proof:}
Let $G=S_t[H,k,j]$ be a thin spider where $H$ is the subgraph induced by the head of the spider, having some vertex ($j>0$).
By the definition of spider we have $k\geq2$.
Using the same labeling that in the statement above, the Laplacian matrix of $S_t$ can be obtaining just replacing
the block corresponding to the vertices in the head by $L(H)+k\mathds{I}$,
where $L(H)$ is the Laplacian matrix of the subgraph $H$.

Then, the matrix $L(S_t)$ can be written as a block matrix:
$$L(S_{t})=
\left[
  \begin{array}{ccc}
    [(k+j+1)\mathds{I}-\mathds{J}]_{k,\, k} & -\mathds{I}_{k} & -\mathds{J}_{k,\, j} \\
    -\mathds{I}_{k} & \mathds{I}_{k} & \Theta_{k,\, j} \\
    -\mathds{J}_{j,\, k} & \Theta_{j,\, k} & L(H)+k\mathds{I}_{j} \\
  \end{array}
\right],
$$

As before, setting $p=k+j$ and $q=(k+j)^2+4$, we have that $({p+2-\sqrt{q}})/2$  is an eigenvalue of the graph,
 independent of the spider's head, with multiplicity at least $k-1 \geq 1$.

However, as $k \geq 0$,  $q=(k+j)^2+4$  is not a perfect square, so $({k+j+2-\sqrt{(k+j)^2+4}})/{2} \not \in \mathds{Z}$
and $S_t$ is not $L$-integral.

If the spider is headless, we have already obtained that $(k+2-\sqrt{k^2+4})/2$ is an $L$-eigenvalue and it is never an integer.
\begin{flushright}
$\blacksquare$
\end{flushright}

We can state the following corollary:

\begin{corollary}\label{integral gorda}
If $G$ is a thick spider then $G$ is not $L$-integral.
\end{corollary}

\noindent \textbf{Proof:}
Let $G=S_T[H,k,j]$ be a thick spider, then its complement is $S_t[\overline{H},k,j]$ which is not $L$-integral, by Theorem \ref{integral magra}. Then, as a consequence of Proposition \ref{Mohar} we conclude that $S_T[H,k,j]$ is not $L$-integral.
\begin{flushright}
$\blacksquare$
\end{flushright}

\begin{remark}
In short, if $G$ is a spider graph, thin or thick, with or without head, it is not $L$-integral.
\end{remark}

\section{\large{$P_4$-Sparse Graphs}}

A cograph is a $P_4$-free graph, i.e. a graph that does not contain a path with four
vertices $P_4$ as an induced subgraph.

In \cite{Hoang}, Hoàng introduced the class of $P_4$-sparse graphs, containing the class of cographs:

\begin{defi}
$G$ is $P_4$-sparse graph if every set of five vertices in $G$ induces at most one $P_4$.
\end{defi}

Directly from the definition we note that, if a $P_4$-sparse graph is disconnected,
then all its connected components are $P_4$-sparse graphs. Also the union of $P_4$-sparses graphs maintain this property.

\begin{remark} \label{complement_p4esp}
 The complement of a $P_4$-sparse graph is also a $P_4$-sparse graph.\\ In fact, supose that $G$ isn't a $P_4$-sparse graph, then there is a set $\{a,b,c,d,e\}\subset V(G)$ such that $A=\{a,b,c,d\}$ and $B=\{a,b,c,e\}$ induce $P_4$, but $P_4 \approx \overline{P_4}$, then $A$ and $B$ induce $P_4$ in $\overline{G}$, ie, this isn't a $P_4$-sparse graph.
\end{remark}

In \cite{Jamison} is showed that a spider is $P_4$-sparse if and only if the subgraph induced by its head is $P_4$-sparse. From this we can see that the graphs in example \ref{exem_aranha} are $P_4$-sparse. It is also proved an important result relating $P_4$-sparse graphs and spider graphs:
\begin{theorem} \label{p4 e aranha}\cite{Jamison}
If $G$ is a non trivial $P_4$-sparse graph, then either $G$ or $\overline{G}$ is disconnected, or $G$ is a spider whose head, if exists, induces a $P_4$-sparse graph.
\end{theorem}

Now, we present the result concerning the L-integrality:
\begin{theorem} \label{p4integral}
Let $G$ be a $P_4$-sparse graph. Then, $G$ is L-integral if and only if $G$ is a cograph.
\end{theorem}

\noindent \textbf{Proof:}

Let $G$ be a non-cograph $P_4$-sparse graph. By Theorem \ref{p4 e aranha} we have three cases to consider:

\begin{enumerate}

  \item $G$ is a spider: by Theorem \ref{integral magra} and Corollary \ref{integral gorda}, $G$ is not $L$-integral and the desired is proved.

  \item $G$ is a disconnected graph: as $G$ is not a cograph then it has a connected component $H$ such that $H$ induces a path $P_4$, ie, $P_4 \subseteq H \subseteq G$. Note that $H$ is a connected $P_4$-sparse then, again by  Theorem \ref{p4 e aranha},  $\overline{H}$ is disconnected or $H$ is a spider. If the latter case occurs, then $H$ is not L-integral and therefore, by Remark \ref{remark_l-espectral}, $G$ is not L-integral, which completes the proof. Otherwise $\overline{H}$ is disconnected, but $\overline{P_4}\thickapprox P_4$ is an induced subgraph of $\overline{H}$, then there is a connected component $H_1 \subseteq \overline{H}$ that induces a $P_4$ and is $P_4$-sparse, by Remark \ref{complement_p4esp}.

      Again, by Theorem \ref{p4 e aranha}, we ensure that $H_1$ is a spider or $\overline{H_1}$ is disconnected.

      If it is a spider, then $H_1$ is not L-integral and so neither $\overline{H}$ and Proposition   \ref{Mohar} ensures that $G$ is not L-integral. On the other hand we have $\overline{H_1}$ disconnected with $\overline{P_4} \thickapprox P_4$ induced subgraph in some connected component $H_2$ of $\overline{H_1}$, ie, $H_2$ is connected $P_4$-sparse. Then, by Theorem  \ref{p4 e aranha}, $H_2$ is a spider or $\overline{H_2}$ is disconnected and  we repeat the procedure.

      Repeating the above procedure we find a spider graph, in a connected component, or a path $P_4$ (which is also a spider). Note that it's not possible to obtain a estable set, since $G$ isn't a cograph. Hence, we have a connected component not L-integral, property that will be transmitted to the original graph $G$.

      \item $\overline{G}$ is disconnected: by Remark \ref{complement_p4esp}, $\overline{G}$ is also $P_4$-sparse. As $G$ induces some $P_4$, we have that $\overline{G}$ induces $\overline{P_4} \thickapprox P_4$, and then $\overline{G}$ is not a cograph. Therefore, $\overline{G}$ satisfies the previous case, concluding that $\overline{G}$ is not L-integral neither is $G$.

\end{enumerate}

\begin{flushright}
    $\blacksquare$
\end{flushright}

We have seen that although any cograph is $L$-integral, a $P_4$-sparse graph non-cograph, is never $L$-integral. And those graphs that have "many" $P_4's$, or simply,  those who every five vertices induce more than one $P_4$, what can we say about the integrality of their $L$-spectrum?
Observing some examples, we see that we cannot conclude anything, as there are examples of this family that are $L$-integral and others that are not. Namely:

\begin{exmp}
$C_6$ and $G$ (in Figure \ref{fig:c6_G}) induce for every five vertices, two $P_4$'s, however the first is $L$-integral  and the second is not.

\begin{figure}[!h]
\centering
\includegraphics[scale=0.3]{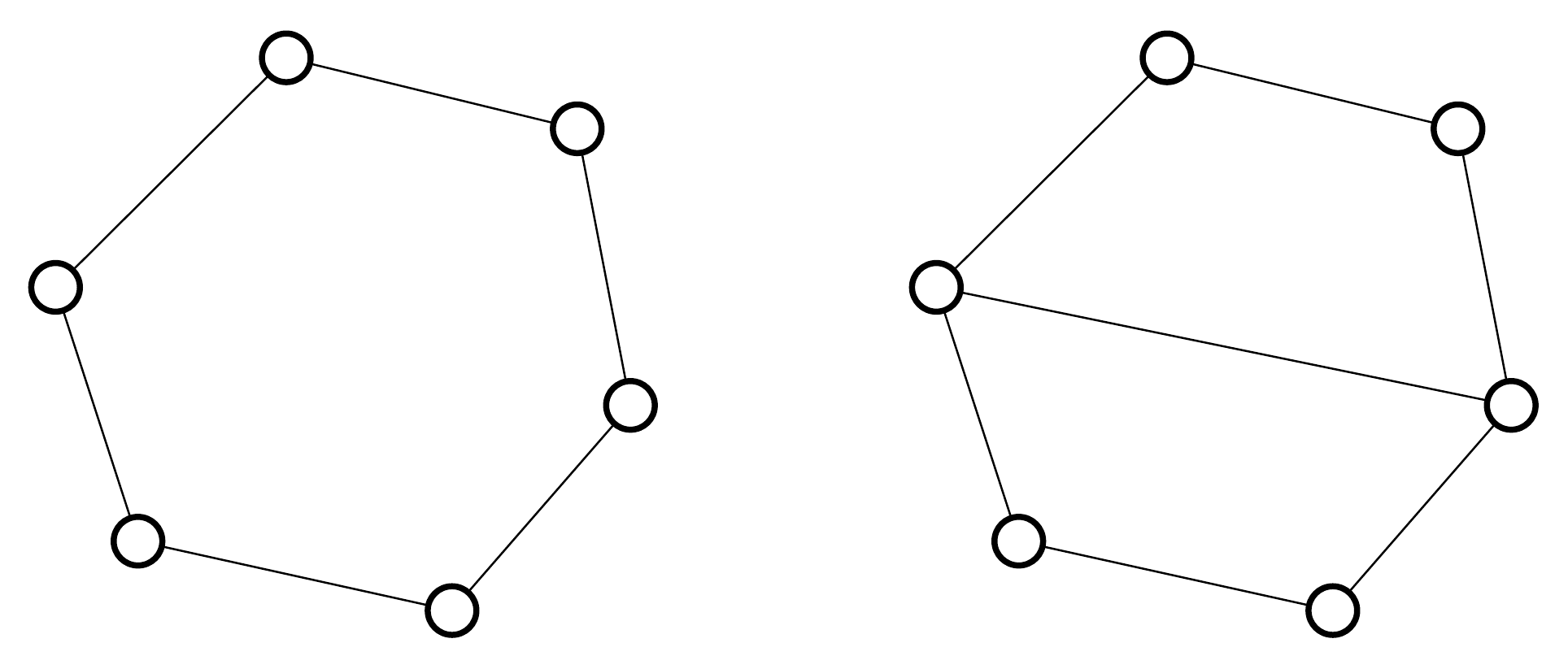}
\caption{$C_6$ e $G$}
\label{fig:c6_G}
\end{figure}

$$\xi(C_6)=\left(
             \begin{array}{cccc}
               4 & 3 & 1 & 0 \\
               1 & 2 & 2 & 1 \\
             \end{array}
           \right)
\quad
\xi(G)=\left(
             \begin{array}{cccccc}
               1 & -1 & -1\pm\sqrt{2} & 1\pm\sqrt{2} \\
               1 & 1 & 1,1 & 1,1 \\
             \end{array}
           \right)$$

\end{exmp}

\subsection{$(q,q-4)-graphs$}

There are other families of graphs characterized by their $P_4$-structure. In the previous section, we present the $P_4$-sparse graphs. Babel and Olariu in \cite{olariu} propose a new class, generalizing the $P_4$-sparse graphs, the class of $(q,t)$ which are those graphs that each $q$ vertices induce at most $t$ $P_4$'s. By this definition we can see that $P_4$-sparse graphs are $(5,1)$ and cographs are $(4,0)$.

Babel and Olariu, enunciate a theorem characterizing a new class, using the following definition:

\begin{defi}
$G(V,E)$ is called $p_4$-connected\footnote{In \cite{olariu}, $p_4$-connected is denoted simply by $p$-connected, but we prefer to emphasize the dependence of the $P_4$-structure.} if, for every partition of $V$ in two sets $A$ and $B$, there is a $P_4$ induced with vertices in $A$ and in $B$.
\end{defi}

\begin{theorem} \cite{olariu} \label{thm_olariu}
Let $G(V,E)$ be a graph $(7,3)$ $p_4$-connected.  Then $|V|<7$ or $G$ is a headless spider.
\end{theorem}

By Theorem \ref{thm_olariu} and Theorem \ref{integral magra}, we conclude the next corollary:
\begin{corollary}
If $G$ is a $(7,3)$ $p_4$-connected graph with at least $7$ vertices, then $G$ is not $L$-integral.
\end{corollary}

\section{\large{$P_4$-extendible graphs}}
The next class was introduced by Jamison and Olariu, in \cite{Jamison2}.
\begin{defi}
A graph $G(V,E)$ is called \textbf{$P_4$-extendible} if,  for any $W\subset V$ inducing $P_4$, there is at most one vertex $x \notin W$ that induces $P_4$ with some vertices of $W$.
\end{defi}

\begin{remark}.
The class of $P_4$-extendible contains strictly the class of cographs and it is distinct from the class of $P_4$-sparse graphs.
\end{remark}

\begin{exmp}
\begin{figure}
\centering
\includegraphics[scale=0.4]{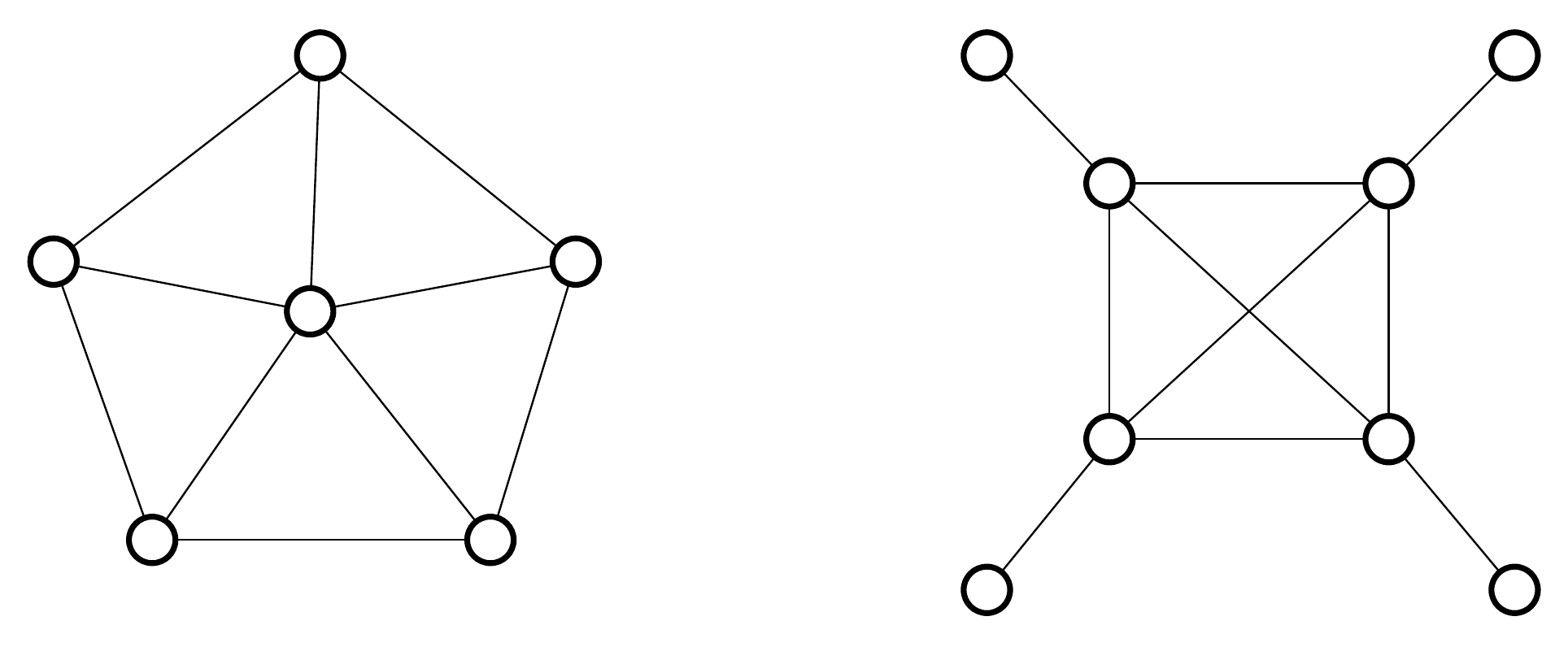}
\caption{$P_4$-extendible not $P_4$-sparse  and  $P_4$-sparse not $P_4$-extendible}
\label{fig:EXt}
\end{figure}
\end{exmp}

In a $P_4$, the vertices of degree $1$ are called \emph{endpoints} and the others are called \emph{midpoints}.
A vertex in $G$ is called an endpoint if it is an endpoint for any  induced $P_4$ in the graph. A  vertex in $G$ is said a midpoint if it is a midpoint for any induced $P_4$ in the graph.

Let us now consider the graphs below

\begin{figure}[!h]
\centering
\includegraphics[scale=0.6]{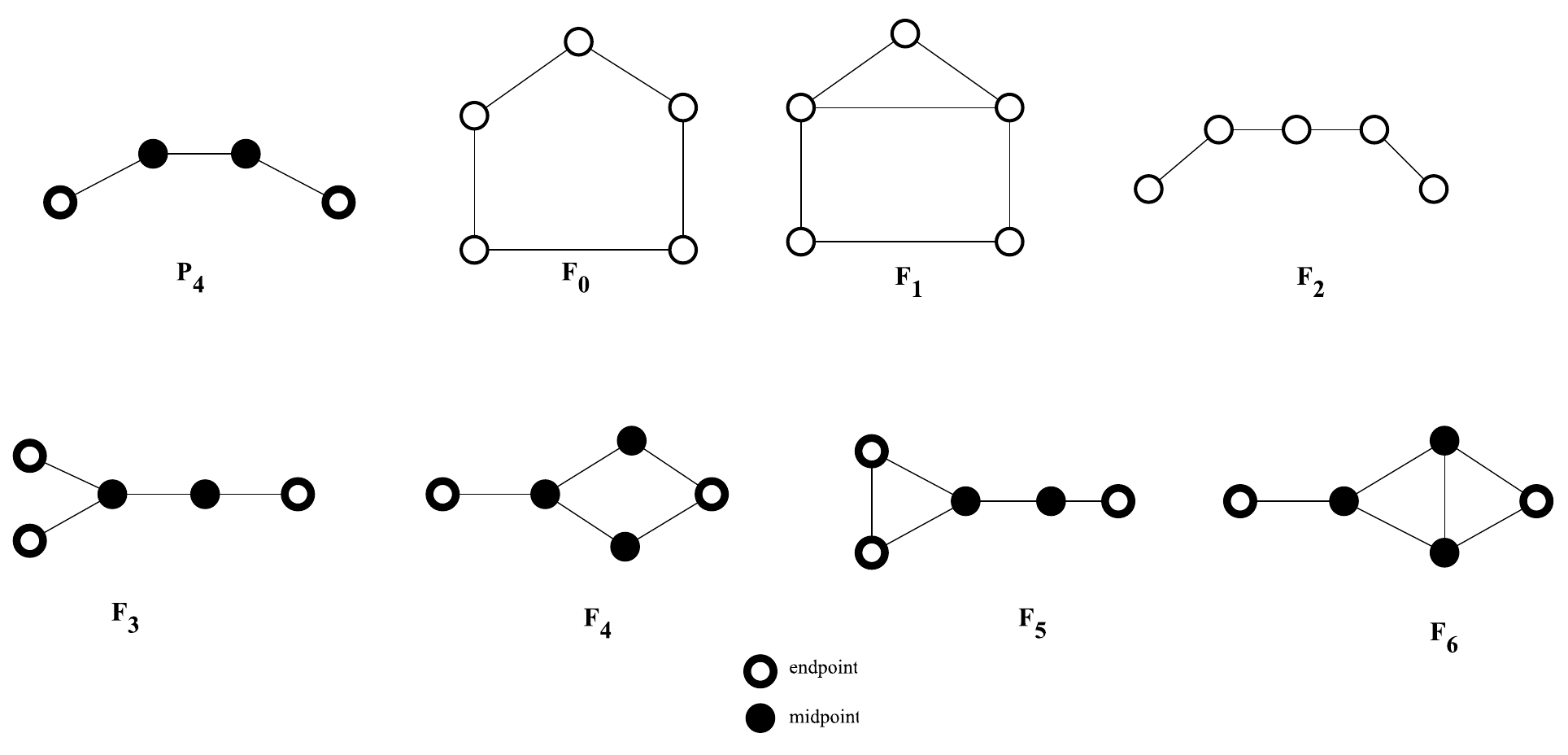}
\caption{$P_4$;$F_0$;$ F_1$;$F_2$;$F_3$;$F_4$;$F_5$;$F_6$}
\label{fig:$P_4$;$F_0$;$ F_1$;$F_2$;$F_3$;$F_4$;$F_5$;$F_6$}
\end{figure}

In \cite{Jamison2}, it is given a characterization of $P_4$-extendible graphs, which will be useful later.

\begin{theorem} \cite{Jamison2} \label{extendible_thm}
If $G$ is $P_4$-extendible with more than one vertex, then it must satisfy exactly one of the conditions below.\\
\textbf{(i)} $G$ is disconnected;\\
\textbf{(ii)} $\overline{G}$ is disconnected;\\
\textbf{(iii)} $G \in \mathcal{F}\cup\{P_4\}$;\\
\textbf{(iv)} there is a subset $D\subset V$ inducing a graph of the set $\{P_4,F_3,F_4,F_5,F_6\}$ and moreover, every vertex in $V(G)\setminus D$  is adjacent to the intermediate vertices and is not adjacent to the extreme vertices of $D$.
\end{theorem}

From the above theorem we conclude that, if $G$ is a connected  $P_4$ -extendible  whose complement is also connected, then $G \in \mathcal{F}\cup\{P_4\}$  (case iii) or $G$ is as illustrated in the figure below.
\begin{figure}[!h]
\centering
\includegraphics[scale=0.6]{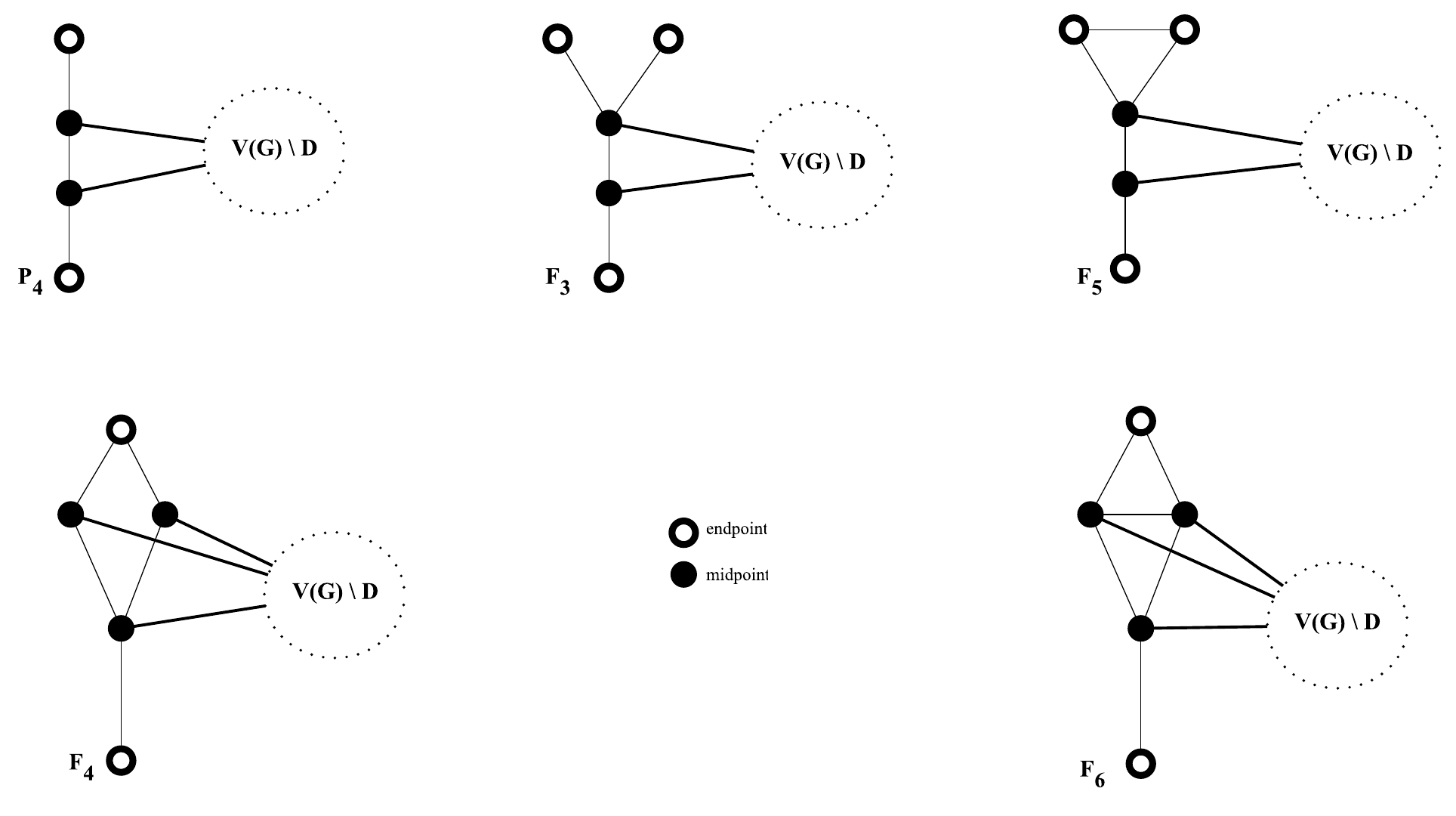}
\caption{case (iv)}
\label{fig:case (iv)}
\end{figure}

We can easily verify that:

\begin{lem}\label{F}
The graphs in $\mathcal{F}\cup\{P_4\}$ are not L-integrals.
\end{lem}
\noindent \textbf{Proof:}
 $P_4$ is not L-integral (it is the headless spider $S_m[2]$). Note that $\overline{F_0}=F_0$, $\overline{F_1}=F_2$, $\overline{F_3}=F_6$, $\overline{F_5}=F_4$. It is therefore sufficient to check that $F_0$, $F_1$, $F_3$ e $F_5$ are not L-integral.

Moreover we have the following Lemma:

\begin{lem} \label{iv}
If $G$ is a graph satisfying the assertion (iv) of theorem \ref{extendible_thm}, then $G$ is not L-integral.
\end{lem}
\noindent \textbf{Proof:}

Let $G(V,E)$ be a graph such that there is $D\subset V$ inducing a graph of $\{P_4,F_3,F_4,F_5,F_6\}$ and every vertice in $V\setminus D$ is adjacent to mid points  of $D$ and not to its endpoints. Let $H$ be the subgraph induced by $V\setminus D$ in $G$, considering $|V(H)|=j\geq 1$. We have five cases to consider:\\
Case 1) $G(D)\approx P_4$: in this case $G \approx S_t[H,4,j]$ and, by theorem \ref{integral magra} $G$ is not L-integral.\\
Case 2) $G(D)\approx F_3$:

We want to determine $x,y,z,w \in \mathds{R}$ such that the vector $\textbf{w}=\left[\begin{array}{ccccccccc}
x & y & 1 & 1 & z & w & w & \ldots & w
\end{array}\right]^t \in \mathds{R}^{j+5}$ is an eigenvector of $L(G)$. This is equivalent to determine $\lambda$ satisfying equality

\begin{small}
$$\left[\begin{array}{ccccccccc}
3+j & -1 & -1 & -1 & 0 & -1 & -1 & \ldots & -1 \\
-1 & 2+j & 0 & 0 & -1 & -1 & -1 & \ldots & -1 \\
-1 & 0 & 1 & 0 & 0 & 0 & 0 & \ldots & 0 \\
-1 & 0 & 0 & 1 & 0  & 0 & 0 & \ldots & 0 \\
0 & -1 & 0 & 0 & 1 & 0 & 0 & \ldots & 0 \\
-1 & -1 & 0 & 0 & 0  &   &   &   &   \\
-1 & -1 & 0 & 0 & 0  &   & L(H)+2\mathds{I} &  &   \\
\vdots & \vdots & \vdots & \vdots & \vdots  &   &   &   &   \\
-1 & -1 & 0 & 0 & 0 &   &   &   &
\end{array}\right]
\left[\begin{array}{c}
x \\ y \\ 1\\ 1\\ z\\ w\\ w\\ \vdots\\ w\\
\end{array}\right]=
\lambda.\left[\begin{array}{c}
x \\ y \\ 1\\ 1\\ z\\ w\\ w\\ \vdots\\ w\\
\end{array}\right]
.$$
\end{small}
The system $S$ can be rewritten as
\begin{equation}
\left\{\begin{array}{c}
\lambda x=3x+jx-y-2-wj\\
\lambda y=-x+2y+jy-z-wj \\
\lambda=1-x \\
\lambda z=z-y \\
\lambda w=-x -y+2w
\end{array}\right. \Rightarrow
\left\{\begin{array}{cc}
x-x^2=3x+jx-y-2-wj & (1)\\
y-xy=-x+2y+jy-z-wj & (2)\\
\lambda=1-x & (3)\\
y=xz & (4)\\
w-xw=-x -y+2w & (5)
\end{array}\right.
\end{equation}

We guarantee that $x\neq 0$, otherwise $y=x=0$ and $\lambda=1$, which generates contradictory values for $w$. Replacing $z=\frac{y}{x}$ of $(4)$, in $(2)-(1)$ we have $y(1-2x-jx-x^2)=-3x^2-jx^2+2x-x^3$.

If $(1-2x-jx-x^2)=0$ then $-3x^2-jx^2+2x-x^3=0$ and, as $x\neq 0$ we have $-x^2-(3+j)x+2=0$ which has no real complex roots, then $\lambda=1-x$ is a non real complex number, which is absurd, since the matrix is symmetric. Then $$y=\frac{-3x^2-jx^2+2x-x^3}{1-2x-jx-x^2} \hbox{ } (6).$$

On the other hand, if $x=-1$ we have $y=1$, $z=-1$ and $\lambda=2$ which generates contradictory values for $w$. Then $x\neq -1$ and by $(5)$ we have $w=\frac{x+y}{x+1}$. So, by $(1)$ and $(4)$ we ensure that $y(x+1+j)=x^3+3x^2+jx^2-2$.

If $x+1+j=0$ we have $x=-1-j$ and $x^3+3x^2+jx^2-2=0$, hence $2j(j+2)=0$ with roots $0$ and $-2$, which generates an absurd, as $j \geq 1$. Therefore we can write
$$y=\frac{x^3+3x^2+jx^2-2}{x+1+j}  \hbox{ } (7).$$

From $(6)$ and $(7)$ we obtain the equation
$$x^5+(2j+4)x^4+(j^2+3j+1)x^3+(-j^2-5j-6)x^2-2x+2=0$$

Note that $x=1$ is a root, hence $y=z=w=1$ and $\lambda=0$, then $\textbf{w}=\vec{\textbf{1}}_{j+5}$, what was expected for the Laplacian matrix. Then we can write $(x-1)q(x)=0$, where

$$q(x)=x^4+(2j+5)x^3+(j^2+5j+6)x^2-2$$

As $j\geq 1$,  $q(0)=-2<0$ and $q(1)=j^2+7j+10>0$. By the Intermediate Value Theorem, the polynomial $q$ has a root in the interval $(0,1)$, then $x \notin \mathds{Z}$. As $\lambda=1-x$, we know that this root should be an irrational number, say $x=\epsilon$.

Then the system solved above have the solution:
{\small $$x=\epsilon \quad y=\frac{\epsilon^3+3\epsilon^2+j\epsilon^2-2}{\epsilon+1+j} \quad z=\frac{\epsilon^3+3\epsilon^2+j\epsilon^2-2}{\epsilon(\epsilon+1+j)} \hbox{ and } w=\frac{\epsilon+\frac{\epsilon^3+3\epsilon^2+j\epsilon^2-2}{\epsilon+1+j}}{\epsilon+1}$$}.

Then $\lambda=1-\epsilon$ is irrational and the graph $G$ is not L-integral.\\

Case 3) $G(D)\approx F_5$: In this case, the Laplacian matrix of $G$ is as the precedent one, only changing the firs block of the matrix for the following:\\
\begin{small}
$$\begin{array}{cccccc}
3+n & -1 & -1 & -1 & 0 & |  \\
-1 & 2+n & 0 & 0 & -1 & |  \\
-1 & 0 & 2 & -1 & 0 & | \\
-1 & 0 & -1 & 2 & 0 & |  \\
0 & -1 & 0 & 0 & 1 & | \\
--- & --- & --- & --- & --- & |  \\
\end{array}$$
\end{small}

Similarly, we want to determine $x,y,z,w,\lambda \in \mathds{R}$ such a way the vector $\textbf{w}$, as taken in case 2, satisfies $L(G).\textbf{w}=\lambda.\textbf{w}$. As this equality leads  to the same system $S$, again $G$ is not L-integral.\\
Case 4) $G(D)\approx F_4$: As $\overline{F_4}=F_5$, $\overline{G}$ is a graph of the previous case. Then, for the proposition \ref{Mohar}, $G$ is not L-integral.\\
Case 5) $G(D)\approx F_6$: Again, we note that $\overline{G}$ is a graph of case 2, as $\overline{F_6}=F_3$ and so $G$ is not L-integral.

\begin{flushright}
$\blacksquare$
\end{flushright}

\begin{remark}
It is easy to check that, if $G$ is $P_4$-extendible then $\overline{G}$ is also $P_4$-extendible. We also have that $G_1$ and $G_2$ are $P_4$-extendibles if and only if  $G_1 \cup G_2$ is $P_4$-extendible.\\
\end{remark}

These two lemmas, along with the above remark, allow us to completely characterize the L-integral graphs in the class of $P_4$-extendible graphs. Again, these are exactly the cographs.

\begin{theorem}
Let $G$ be a $P_4$-extendible graph. Then, $G$ is L-integral if and only if $G$ is a cograph.
\end{theorem}

\noindent \textbf{Proof:}
We will apply Theorem \ref{extendible_thm} in $G$ and, making use of the remark above, we look for a connected component with connected complement that, in turn, satisfies (iii) or (iv) of Theorem \ref{extendible_thm}, and so it is not L-integral by the Lemma \ref{F} and Lemma \ref{iv} .

\begin{flushright}
$\blacksquare$
\end{flushright}

There are other classes of graphs as $P_4$-reducible graphs,\cite{Jamison1} and $P_4$-lite graphs, \cite{Jamison3}, composed by graphs containing a restricted number of $P_4$. The class of $P_4$-reducible graphs is the intersection of $P_4$-sparse and $P_4$-extendible graphs and obviously contain the cographs. So, a $P_4$-reducible graph is $L$-integral if and only if it is a cograph.
It seems that the $L$-integrality is related to the $P_4$ structure of the graph. In this paper we have analyzed
the behavior of the spectrum, related to L-integrality, for some classes.
It remains to search for L-integral graphs in other classes, as $P_4$-lite graphs and $(q,q-4)$-graphs.

\section*{Acknowledgements}
First author was partially supported by CNPq- Grant 476363/2012-8 and CAPES-Grant 99999.002658/2015-01. Second author was partially supported by CAPES.

\end{document}